\documentclass[sigconf,10pt, nonacm]{acmart}

\setcopyright{none}
\renewcommand\footnotetextcopyrightpermission[1]{}
\usepackage{amsmath}
\usepackage{graphicx}
\usepackage{tikz-cd}
\usepackage[dvipsnames]{xcolor}
\usepackage{xspace}
\usepackage{xparse}
\usepackage{subcaption}
\usepackage[normalem]{ulem}
\usepackage{cancel}

\newif\ifshownotes
\shownotestrue

\NewDocumentCommand{\parabf}{s o m}{%
  \IfBooleanTF{#1}{%
    \IfNoValueTF{#2}{\paragraph*{\textbf{#3}}}{\paragraph*[#2]{\textbf{#3}}}%
  }{%
    \IfNoValueTF{#2}{\paragraph{\textbf{#3}}}{\paragraph[#2]{\textbf{#3}}}%
  }%
}

\graphicspath{{figures/}}

\begin{document}

\title{ML-for-ML}

\author{Yutong Zhao}
\authornote{Equal contribution.}
\affiliation{%
  \institution{UCL and BUPT}
  \country{United Kingdom}
}

\author{Noga H. Rotman}
\authornotemark[1]
\affiliation{%
  \institution{UCL}
  \country{United Kingdom}
}

\author{Gianni Antichi}
\affiliation{%
  \institution{Politecnico di Milano}
  \country{Italy}
}

\author{Ran Ben Basat}
\affiliation{%
  \institution{UCL and Broadcom}
  \country{United Kingdom}
}

\begin{abstract}

AI training workloads are growing rapidly, making their time, energy, and infrastructure costs increasingly important. In shared cloud clusters, training and fine-tuning jobs compete with co-running workloads for network resources, while network mechanisms and ML training choices are typically optimized separately: networking controls how bytes move, whereas ML systems control when and how much communication occurs. We argue that this separation leaves end-to-end performance \mbox{on the table.}

We present ML-for-ML, a cross-layer perspective in which network-side and ML-side knobs are selected jointly under a shared time-to-target-loss objective.
Our preliminary prototype shows that by co-optimizing the ML and network parameters, we reach the target loss up \mbox{to 42\% faster.}
\end{abstract}

\maketitle

\section{Introduction}
\label{sec:introduction}

Many distributed machine learning (ML) training jobs run in shared cloud environments~\cite{liberty2020sagemaker}, academic clusters~\cite{xu2025sing}, or multi-tenant datacenter environments~\cite{jeon2019multitenant}, where the same network fabric carries traffic from co-running workloads and background services.
In such environments, ML training traffic competes with other workloads for network bandwidth~\cite{gebara2021network,cao2024crux}, and network conditions can change throughout training~\cite{viswanathan2020mlfabric}. 
Adapting to this variability by deciding when, how much, and how to communicate is, therefore, critical to achieving \mbox{high training performance.}

Traditionally, this problem has been addressed primarily from the networking perspective.
Congestion control~\cite{zhu2015congestion, araujo2026resilient}, network-aware scheduling~\cite{rajasekaran2024cassini}, and routing/topology optimization~\cite{al2008scalable} adapt the network to accommodate the communication demands of distributed training, mitigating contention and improving utilization and training performance.
These mechanisms react to changing network conditions but treat the communication workload as \mbox{an external input.}

However, the communication workload is itself pro\-gram\-ma\-ble. 
In addition to statistical efficiency, ML-side choices such as batch size~\cite{goyal2017accurate,mccandlish2018empirical}, gradient accumulation~\cite{narayanan2021efficient}, and local steps~\cite{mcmahan2017communication,stich2019local} affect when or how much data is exchanged across the network. 
Rather than relying solely on network-side mechanisms to react to changing load, we argue that they should be selected jointly with ML-side decisions. 
The objective is ultimately shared: reaching a target model quality while minimizing costs such as execution time or energy. 
Improvements on one side naturally create opportunities on the other. 
For example, a less contended network may make more frequent synchronization preferable, while lowering synchronization frequency or batching decisions can directly \mbox{reduce network contention.}

\begin{figure}[t]
    \centering
    \includegraphics[width=\linewidth]{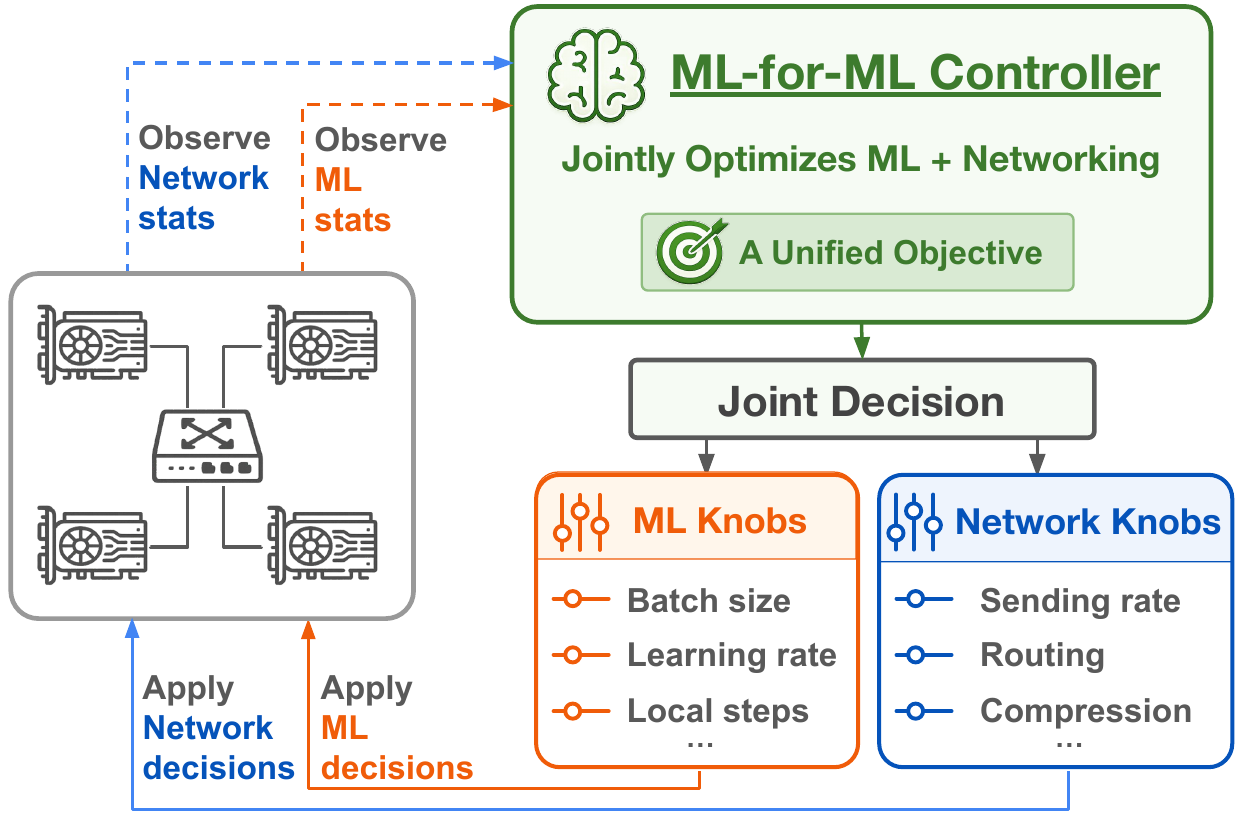}
    \caption{The envisioned ML-for-\xcancel{(\sout{Networking-for}})-ML 
    loop: the controller observes training progress and network conditions, jointly selects ML-side and network-side knobs under a unified objective, and applies the resulting decisions to the training job and the network. 
    Updated ML and network measurements then inform the next decision, \mbox{closing the loop.}
    }
    \label{fig:ml4ml-loop}
\end{figure}

The problem lies at the intersection of two active but largely separate communities.
ML-for-networking uses ML to improve network mechanisms such as congestion control~\cite{winstein2013remy,jay2019aurora} and traffic engineering~\cite{valadarsky2017learning, perry2023dote}.
Networking-for-ML improves transport and rate control~\cite{zhu2015congestion,trimmablegradients}, in-network aggregation~\cite{gebara2021network,THC}, and network-aware communication and job scheduling~\cite{rajasekaran2024cassini,cao2024crux} for training workloads.
We envision a unified loop, as illustrated in Figure~\ref{fig:ml4ml-loop}, in which the training system observes its progress and network conditions and optimizes network-side and ML-side knobs jointly under the \mbox{shared control objective.}

Realizing this loop introduces several challenges.
First, the usefulness of a knob combination depends on both training progress and network load.
Second, decisions must be made online, as the future background traffic load is unknown.
Finally, estimating the training progress (e.g., validation loss) or network conditions (e.g., background load) requires resources such as running a validation set or injecting telemetry probes, which compete with the \mbox{job's own resources.}

As a first step toward this vision, this paper investigates whether optimizing network-side and ML-side knobs independently is sufficient.
Our methodology evaluates candidate combinations by asking how each combination changes both foreground-visible communication time and training progress. 
Our case study instantiates this with two representative knobs: communication volume on the network side and batch sizing on the ML side. 
Across the evaluated background-load traces, we show that composing independently selected knob settings 
takes 1.13--1.42\(\times\) the time required by joint optimization to reach \mbox{the target quality.}

More broadly, as distributed training scales and diversifies, we envision broadening the scope of ML-for-ML to include richer joint action spaces, broader objectives and workloads, other deployment settings, and coordination among multiple adaptive jobs.
As the networking-for-ML and ML-for-networking communities mature, we call for mechanisms and interfaces that unite the communities around \mbox{shared end-to-end goals.}

\section{ML-for-ML}
\label{sec:methodology}

This section defines the general joint optimization problem over network-side and ML-side knobs underlying the ML-for-ML loop.
The controller's action space consists of combinations of network-side and ML-side knobs.
Table~\ref{tab:knobs} lists \mbox{its representative examples.}

The preferred knob combination depends on the current training phase, the network's background load, and interactions among the knobs. 
Because these conditions change over time and differ across workloads, a combination selected before training need not remain preferable throughout the run. 
A learned controller provides a natural way to address this state-dependent decision problem. 
It can learn from prior observations and periodically update the joint choice during training, rather than optimize the network-side and ML-side \mbox{knobs in isolation.}

The remainder of this section characterizes the joint decision structure that such a controller must capture.
We state the control objective, analyze why independent knob tuning can fail, describe what the controller should observe, and explain how those observations \mbox{lead to joint optimization.}

\begin{table*}[t]
\centering
\caption{
ML-side and network-side control knobs, 
with representative works highlighting their impact on distributed training performance}
\label{tab:knobs}
\begin{tabular}{c|p{0.68\linewidth}}
\hline
Type & Possible knobs \\
\hline
ML & 
Batch size~\cite{goyal2017accurate,mccandlish2018empirical}; 
gradient accumulation~\cite{narayanan2021efficient}; 
learning rate~\cite{smith2017cyclical, loshchilov2017sgdr}; 
local steps~\cite{mcmahan2017communication,stich2019local}; 
worker synchronization / participation~\cite{dean2012large, ho2013more} 
\\
\hline
Network & 
Sending rate~\cite{zhu2015congestion, araujo2026resilient};
routing~\cite{al2008scalable}; 
communication precision/compression~\cite{seide20141, aji2017sparse,han2024beyond,han2026dynamiq}; 
redundancy / reliability trade-off~\cite{vulimiri2013low};
bandwidth allocation / prioritization~\cite{alizadeh2013pfabric} 
\\
\hline
\end{tabular}
\end{table*}

\subsection{Control Objective}
\label{sec:method-tta}

Although our framework can optimize for other objectives, such as minimizing energy consumption, for ease of exposition, we hereafter consider the time-to-target loss goal: minimizing the wall-clock time for the foreground model to reach a \mbox{fixed evaluation loss.}

This end-to-end objective depends jointly on training progress,
measured by evaluation loss, and the wall-clock cost of training,
including both local computation and network communication.
We therefore assess progress toward the target loss as a function
of \mbox{elapsed wall-clock time.}

\subsection{Why Independent Knob Optimization Can Fail}
\label{sec:method-nonseparable}

A natural baseline is to optimize one knob at a time, holding all other knobs at their baseline settings, and then compose the resulting choices.
However, this combination need not be the best choice over the full joint action space because network-side and ML-side knobs are \mbox{generally not separable.}

A knob choice can change communication volume, communication timing, update frequency, or training dynamics, thereby altering the effects of the other knobs.
A value that appears preferable when evaluated at the baseline settings can therefore become suboptimal once combined with other independently selected values.
Conversely, the best joint combination can include a value that does not appear best \mbox{when evaluated alone.}

For example, communication compression can reduce communication cost but may also alter training progress.
Batch-size choices can change how often the model is updated and how often workers communicate, as well as the resulting loss trajectory.
These effects are coupled: compression can provide greater benefit when communication is frequent, but less additional benefit under a batch-size setting that already communicates less often.
Conversely, once compression has reduced communication cost, a batch-size setting that further reduces communication frequency may no longer offer the best tradeoff between communication cost and training progress.
Thus, the value of each setting depends on the other, and the settings preferred when the knobs are evaluated separately need not form the best combination. 
This motivates evaluating network-side and ML-side knobs jointly, rather than composing independently \mbox{selected single-knob settings.}

\subsection{Observing the Training Process and Network Conditions}
\label{sec:method-pressure}

Choosing a combination setting of knobs 
requires observations from both sides of the system.
The controller may draw on both raw system signals and prior knowledge.
Network-side inputs may include recent collective durations, active probes, NIC counters, switch telemetry, queueing depth, ECN marks, or packet-level delay.
ML-side inputs may include loss trends, step times, sample efficiency, or \mbox{known loss costs.}

Network conditions and training dynamics, including background traffic, training phase, and communication placement, change throughout a run, so a one-time profile may quickly become outdated.
A deployed controller must therefore refresh its observations and make decisions periodically over short intervals. 
The duration of each interval is constrained by the time required to collect the relevant network and loss telemetry and adapt the knobs.
It must also account for the delay between observation and actuation, as conditions may change before a selected combination takes effect. 
Rather than reacting only to the latest measurement, the controller can use current and historical observations to anticipate the conditions under which the decision \mbox{will take effect.}

\subsection{From Observations to Joint Optimization}
\label{sec:method-policy-generation}

The controller does not merely need a scalar label such as ``congested.'' 
It needs to translate network and ML signals into estimates that distinguish among candidate combinations, because the same observed state may affect candidate combinations \mbox{in different ways.}

As a practical design choice, we use a greedy per-interval decision rule as a proxy for the global time-to-target objective.
At each decision point, the controller estimates the expected loss reduction over the upcoming interval for every combination of network-side and ML-side knob settings under the anticipated training and network conditions.
These estimates account for both the foreground-visible communication time and the training progress achievable within the interval.
The controller then selects the combination with the largest estimated loss reduction.
Although this local rule is not an exact solution to the global objective, it provides a practical online rule for comparing combinations and updating choices \mbox{as observations change.}

This comparison provides the basis for testing the necessity of joint optimization while accounting for both timing and training effects.
It evaluates whether network-side and ML-side knob combinations lead to different choices from those obtained by composing independently optimized single-knob decisions.
Sensing, prediction, and actuation are needed to implement this rule in a full online controller; the methodology here focuses on the joint comparison that such a \mbox{controller relies on.}

\section{A Two-Knob Case Study}
\label{sec:evaluation}

We instantiate the methodology with a two-knob case study:
one network-side knob that changes communication volume, and one ML-side knob that changes batch sizing. 
This small combination space lets us test the key question directly: 
whether independently chosen knobs can compose into a suboptimal system-level outcome, and whether joint \mbox{optimization avoids it.}

\subsection{Case-Study Setup}
\label{sec:evaluation-setup}

\noindent\textbf{Control knobs.}
The network-side knob is communication precision \(p\), the numeric format of each gradient exchange, which changes communication volume. Here, the action space is \(p\in\{\mathrm{BF16},\mathrm{MXFP4}\}\)~\cite{rouhani2023mx}.
The ML-side knob is the gradient accumulation factor \(g\), the number of microbatches accumulated before a data-parallel (DP) synchronization and optimizer update. 
It changes the batch size and how often the training job synchronizes. 
Here, \(g\in\{4,16\}\), corresponding to GA4 and GA16, respectively. 
Across GA settings, the learning rate follows the \mbox{square-root scaling rule}~\cite{hoffer2017train}. 

\begin{table}
  \centering
  \scriptsize
  \caption{
  Representative policy choices during training under contention generated by \mbox{co-running GPT-1B jobs.}
  }
  \label{tab:gpt1b-bg-corun-policy}
  \setlength{\tabcolsep}{4pt}
  \begin{tabular}{lll}
    \hline
    Policy & Rule & Representative choice 
    \\
    \hline
    \textsc{Static}         & fixed               & GA4-BF16    
    \\
    \textsc{Knob-Precision} & precision only      & GA4-MXFP4   
    \\
    \textsc{Knob-GA}        & GA only             & GA16-BF16   
    \\
    \textsc{Decoupled}      & compose projections & GA16-MXFP4  
    \\
    \textsc{Joint}          & joint optimization & GA4-MXFP4   
    \\
    \hline
  \end{tabular}
\end{table}
\begin{figure*}
\centering
    \begin{subfigure}[t]{0.245\textwidth}
      \centering
      \includegraphics[width=\linewidth]{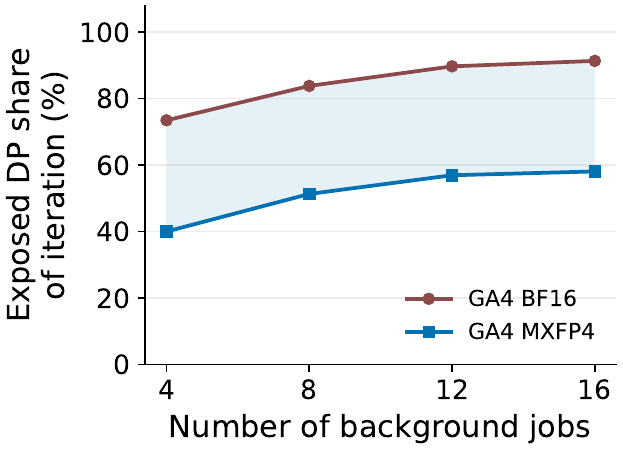}
      \caption{DP exposure}
    \end{subfigure}\hfill
    \begin{subfigure}[t]{0.245\textwidth}
      \centering
      \includegraphics[width=\linewidth]{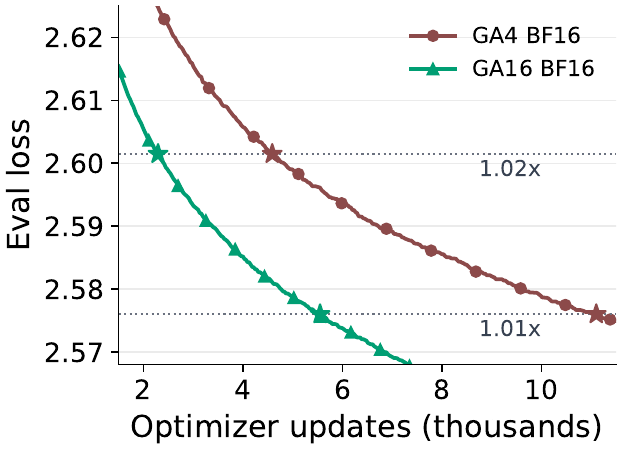}
      \caption{GA tradeoff}
    \end{subfigure}\hfill
    \begin{subfigure}[t]{0.245\textwidth}
      \centering
      \includegraphics[width=\linewidth]{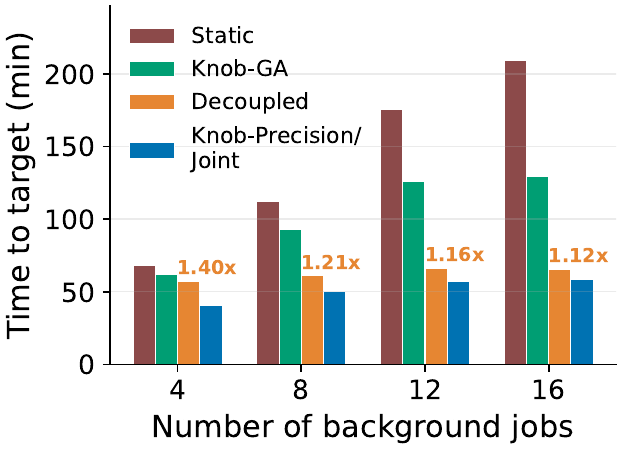}
      \caption{\(1.02\times L_{\min}\)}
    \end{subfigure}\hfill
    \begin{subfigure}[t]{0.245\textwidth}
      \centering
      \includegraphics[width=\linewidth]{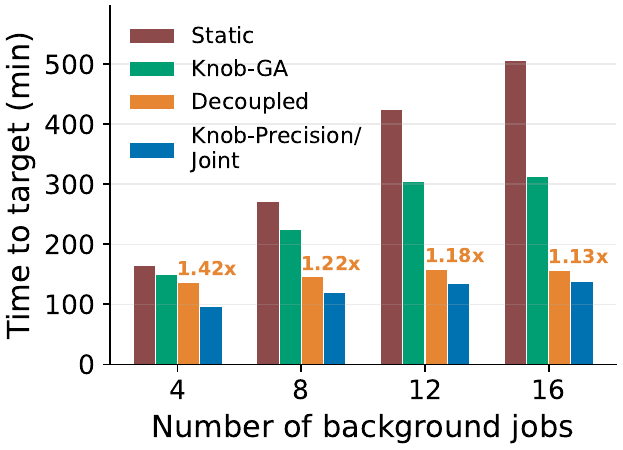}
      \caption{\(1.01\times L_{\min}\)}
    \end{subfigure}

    \caption{Policy behavior under contention generated by co-running GPT-1B jobs.
    Panel (a) shows that, under fixed GA4, compression reduces the exposed DP share of each foreground-job iteration from 73--91\% with BF16 to 40--58\% with MXFP4.
    Panel (b) shows the complementary GA tradeoff under fixed BF16: larger GA reduces communication frequency but changes the number of optimizer updates required to reach a target loss.
    Panels (c) and (d) report end-to-end first-crossing times to \(1.02L_{\min}\) and \(1.01L_{\min}\), respectively, where \(L_{\min}=2.55048\) is the final evaluation loss measured for the \textsc{Static} baseline.
    \textsc{Joint} coincides with \textsc{Knob-Precision} and is shown as the blue bar, while \textsc{Decoupled} composes the two single-knob projections and \mbox{is consistently slower.}
  }
    \label{fig:gpt1b-bg-corun-summary}
\end{figure*}

\noindent\textbf{Foreground workload.}
The foreground task we set is GPT-2 Large~\cite{radford2019language}, a 774M-parameter model with 36 layers, a hidden size of 1280, 20 attention heads, and a sequence \mbox{length of 1024.}

\noindent\textbf{Policies and targets.}
Based on these two knobs, we compare five policies throughout. 
\textsc{Static} fixes the baseline combination at GA4-BF16.
\textsc{Knob-Precision} only varies precision \(p\) while holding the ML-side knob at its baseline setting;
\textsc{Knob-GA} only varies \(g\), which raises the batch size and lowers the synchronization frequency together, while holding the network-side knob at its baseline setting. 
\textsc{Decoupled} optimizes each knob independently and composes
the two resulting values, 
while \textsc{Joint} evaluates over the full combination space. 
We use \(L_{\min}=2.55048\), the final evaluation loss measured for
the \textsc{Static} baseline, as the common reference for the
target-loss thresholds used to \mbox{compare all policies.}

\noindent\textbf{Timing model and measurement setup.}
We obtain wall-clock evaluation-loss trajectories by combining separate measurements of computation, communication, and training progress. 
The foreground iteration time under network condition \(z\) is
modeled as
\[
  T_{\mathrm{iter}}(g,p,z) := gT_c + T_{\mathrm{dp}}(p,z).
\]
We profile \(T_c=59.3\) ms, the average per-microbatch compute time of our foreground job, from forward-backward passes on A100 GPUs.
\(T_{\mathrm{dp}}(p,z)\) denotes the DP time exposed to the foreground job after accounting for overlap with local computation.
We estimate \(T_{\mathrm{dp}}(p,z)\) using SimAI/NS-3 replays~\cite{wang2025simai} driven by co-run traffic traces, with the simulator configured to match the A100-cluster setting.
Furthermore, we obtain the corresponding evaluation-loss trajectories from PyTorch training runs under the knob choices of each policy.
With a faster GPU, \(T_c\) decreases and communication accounts for a larger share of iteration time.
This may shift the controller's preferred knob combination, while the \mbox{formulation remains unchanged.}

We evaluate the policies in two complementary settings: 
one examines the insufficiency of composing the knobs independently, 
while the other examines how the preferred joint choice differs \mbox{across contention conditions.}

\begin{figure*}[t]
  \centering
  \includegraphics[width=0.95\linewidth]{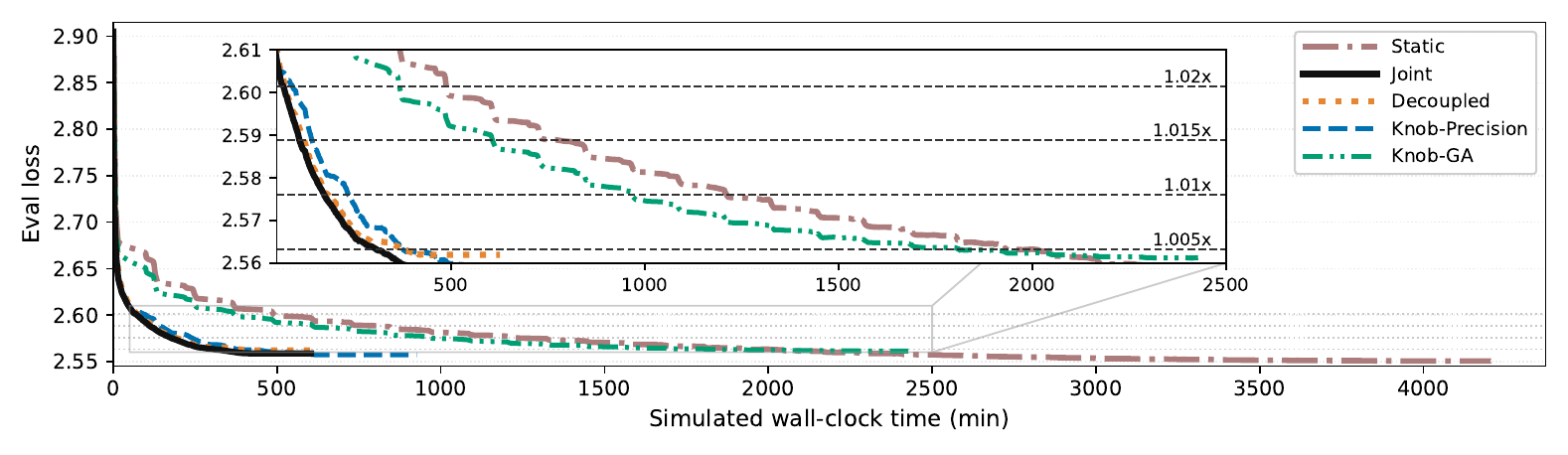}
  \caption{
  Evaluation-loss trajectories under the contention schedule generated by co-running GPT-13B jobs.
  Adaptive policies update their knob settings over the course of the run.
  The inset highlights first-crossing times for the four common target losses, \(1.005\), \(1.01\), \(1.015\), and \(1.02\) \mbox{times \(L_{\min}\).}
  }
  \label{fig:gpt13b-dynamic-scenario}
\end{figure*}

\subsection{The Cost of Decoupled Optimization}
\label{sec:gpt1b-bg-corun}

This scenario compares single knob, decoupled optimization, and joint optimization to assess the end-to-end impact of coordinating the two knobs.
The background traffic is recorded from concurrently running GPT-1B jobs that alternate between non-DP computation and DP communication, share the network, and interact through congestion control. 
The resulting traces preserve the phase-structured and bursty contention produced by co-running training workloads.
We vary the number of background jobs, \(N\in\{4,8,12,16\}\). 
To characterize the resulting network load, we report 95th-percentile background traffic volumes of 17.55, 22.02, 25.83, and 26.16 GB per 50 ms, respectively. 
For each value of N, we replay the foreground job against the corresponding traffic trace at four relative start offsets and average the exposed DP time across the replays, reducing sensitivity \mbox{to phase alignment.}

Table~\ref{tab:gpt1b-bg-corun-policy} captures the five policies' representative choice during training. 
Panels (a) and (b) of Figure~\ref{fig:gpt1b-bg-corun-summary} show the two one-dimensional projections underlying these choices. 
At this point, the precision projection prefers MXFP4 at fixed GA4, while the GA projection prefers GA16 at fixed BF16. 
The resulting choices are \textsc{Knob-Precision} \(\rightarrow\) MXFP4 and \textsc{Knob-GA} \(\rightarrow\) GA16, which \textsc{Decoupled} composes into GA16-MXFP4. 
\textsc{Joint}, by contrast, evaluates all four combinations directly.
Under MXFP4, the remaining exposed DP cost is low enough that the additional communication savings from GA16 do not offset its training-progress cost. 
\textsc{Joint} therefore chooses GA4-MXFP4, matching the precision-only projection but \mbox{differing from \textsc{Decoupled}.}

Panels (c) and (d) report end-to-end first-crossing times for the five policies at \(1.02L_{\min}\) and \(1.01L_{\min}\).
At both targets, \textsc{Decoupled} is slower than \textsc{Joint}, while \textsc{Knob-Precision} matches \textsc{Joint}.
At \(1.01L_{\min}\), the first-crossing-time ratios relative to \textsc{Joint} across \(N\in\{4,8,12,16\}\) are 1.13--1.42\(\times\) for \textsc{Decoupled}, 1.71--3.65\(\times\) for \textsc{Static}, and 1.54--2.26\(\times\) for \textsc{Knob-GA}.
As \(N\) increases, the relative slowdowns of \textsc{Static} and \textsc{Knob-GA} increase with \(N\), because both retain BF16 as communication pressure grows.
By comparison, \textsc{Decoupled}'s absolute first-crossing time increases, but its slowdown relative to \textsc{Joint} narrows from \(1.42\times\) to \(1.13\times\).
As traffic becomes heavier, the value of reducing synchronization frequency increases, \mbox{narrowing this relative gap.}

\subsection{A Moving Optimum}
\label{sec:gpt13b-boundary}

We next examine whether the preferred joint combination differs across contention levels  for the same training job.
This scenario spans a wide range of foreground-visible communication pressure, so a combination that is preferred at one level need not be preferred at another.

We construct contention levels from zero to five concurrently running GPT-13B training jobs.
Each co-running job follows its measured computation and DP-communication cycle, preserving the temporal structure of the training workload.
Jobs arrive at a Poisson rate of \(0.002\,\mathrm{s}^{-1}\) and release the network when they finish, so the number of active jobs varies over the run, averaging 2.7.
All policies experience the same arrival schedule for a fair comparison, and the adaptive policies update their choices as training progresses and foreground-visible \mbox{network conditions change.}

The schedule produces substantial variation in the communication cost exposed to the foreground job.
Compression reduces communication exposure at every contention level, but leaves a larger residual DP cost under stronger contention.
As this residual cost grows, the communication savings from reducing synchronization frequency through a larger GA factor can become large enough for less frequent synchronization \mbox{to be worthwhile.}

This changing tradeoff is reflected in the choices \textsc{Joint} makes.
GA4-MXFP4 remains its most frequent selection at every contention level, but the share of GA16-MXFP4 rises monotonically with the number of background jobs, from 0\% with no background traffic to 40\% at the highest level.
Under light contention, MXFP4 reduces the exposed DP cost enough that the additional communication savings from GA16 do not offset its training-progress cost;
as contention grows, the residual DP cost after compression stays high enough that less frequent synchronization becomes worthwhile for a growing fraction of updates.
\textsc{Decoupled}, by contrast, selects GA16-MXFP4 for 43--58\% of its updates based on the additive combination of the single-knob choices.

Figure~\ref{fig:gpt13b-dynamic-scenario} compares the evaluation-loss trajectories of the five policies.
We report first-crossing times at four common target losses reached by all policies: \(1.005\), \(1.01\), \(1.015\), and \(1.02\) times \(L_{\min}=2.55048\).
Across all four targets, \textsc{Joint} reaches each target first.
Relative to \textsc{Joint}, the slowdown ranges from 5.4\% to 20.8\% for \textsc{Decoupled} and 22.7\% to 36.3\% for \textsc{Knob-Precision}, while \textsc{Knob-GA} and \textsc{Static} are slower by 4.4--4.9\(\times\) and 5.3--6.3\(\times\), respectively.

The two single-knob policies capture complementary benefits:
\textsc{Knob-Precision} reduces communication volume via compression, while \textsc{Knob-GA} reduces synchronization frequency.
Each is useful on its own: across the four targets, \textsc{Knob-Precision} is 5.1--5.6\(\times\) faster than \textsc{Static} and \textsc{Knob-GA} is 1.1--1.3\(\times\) faster.
\textsc{Decoupled} takes both and reaches 5.2--6.8\(\times\), while \textsc{Joint} reaches 6.3--7.3\(\times\) from the same two knobs.
The remaining margin is therefore not a knob that \textsc{Decoupled} lacks: it comes from re-evaluating the pair as contention shifts the residual communication cost after compression, which \mbox{composing two independent projections cannot do.}

\subsection{Takeaways}
Taken together, our initial results show that treating individual knobs independently does not reliably identify the best combination, and that the optimum \mbox{changes with contention.}
By jointly optimizing the network-side and ML-side knob settings, \textsc{Joint} achieves the best observed first-crossing time across all \mbox{reported target losses.}

\section{Discussion}
\label{sec:discussion}

Treating our case study as a starting point, we outline the research directions opened up by this control interface:
extending the ML-for-ML perspective to richer control spaces,
broader objectives and workloads, other deployment settings, and
\mbox{multiple adaptive jobs.}

\parabf{Toward a knob taxonomy for ML-for-ML}
The two knobs studied in this paper cover only a small part of a broader design space. 
Network-facing controls include routing, collective algorithms, compression, and congestion response; 
training-facing controls include gradient accumulation, batch size, checkpointing, optimizer choice, and learning-rate schedules;
resource-facing controls include GPU count, placement, scheduling, and power caps. 
The community needs a set of knobs that should be controlled together, while taking interactions into account, to ensure overall utility.

\parabf{Beyond time: resources, energy, and cost.}
We use time-to-target loss because it is a clean objective shared by networking and ML decisions, but it is not the only one. 
In cloud and multi-tenant settings, users and operators may also care about GPU-hours, dollar cost, energy consumption, power limits, fairness, or meeting deadlines under a fixed budget. 
Once the objective expands, the resource-facing controls above become first-class: GPU allocation, placement, and power caps are direct actions for trading time against cost and energy. 
A broader agenda, therefore, asks not only which action minimizes time-to-target loss but also which action best balances user-
\mbox{and cluster-level objectives.}

\parabf{Beyond datacenter training.}
We focus on distributed training over a shared datacenter network, but the joint-control perspective extends to other settings. 
In federated learning, for example, participating workers differ in bandwidth, latency, compute, and availability, so the relevant system state is participant-specific rather than summarized by a single shared background-load signal. 
The same principle applies, with selection made per participant rather than once for the job, which raises the question of whether effective per-participant decisions can be made without a global \mbox{view of the system.}

\parabf{Across training stages and into serving.}
The current design applies to both pre-training and fine-tuning: both provide an evaluation-loss trajectory against which network-side and ML-side choices can be assessed, so time to a target evaluation loss remains a suitable objective. 
Serving is a related extension in which the objective shifts from loss reduction to outcomes such as latency, throughput, and quality.
ML-side choices such as batching, precision, parallelism, and placement continue to interact with network-side choices such as routing, prioritization, and transport behavior.
The corresponding research question is how to coordinate these choices under changing request workloads and service-level objectives, where loss is no longer \mbox{the feedback signal.}

\parabf{Interaction with other jobs.}
Our experiments study one foreground job adapting to background traffic. 
A shared cluster raises the question of how the system behaves when background jobs adapt too. For example, bandwidth-hungry jobs might increase their sending rates as a direct consequence of our job reducing it.
While we expect ML-for-ML to allow multiple foreground jobs and be a fair network citizen, running alongside inconsiderate or adversarial flows may \mbox{pose interesting challenges.}

\section{Related Work}
\label{sec:related-work}

\parabf{Network-aware optimizations for ML systems.}
Prior work focuses on network-side mechanisms to accommodate the heavy communication demands of distributed training. 
These include transport-layer rate control~\cite{zhu2015congestion,araujo2026resilient}, network-aware job scheduling~\cite{rajasekaran2024cassini}, and routing or topology optimization~\cite{al2008scalable}. Collectively, these approaches optimize network behavior while treating the training process and the communication demand it \mbox{generates as given.}

\parabf{Communication-efficient training optimization.}
A complementary direction adapts the training process to reduce or hide communication overhead. These techniques include gradient quantization and sparsification~\cite{seide20141,aji2017sparse}, local steps~\cite{mcmahan2017communication,stich2019local}, and gradient accumulation~\cite{narayanan2021efficient}. While some approaches jointly optimize internal ML knobs (e.g., batch size and learning rate~\cite{smith2018don}), they do not explicitly model, provide visibility into, or dynamically \mbox{alter network behavior.}

\parabf{Adaptive and joint optimization.}
Closest to our work are runtime optimization frameworks. For instance, systems like KungFu~\cite{mai2020kungfu} dynamically tune training configurations and execution strategies, while Pollux~\cite{qiao2021pollux} co-adapts training configuration and resource allocation to optimize training goodput. Adaptive distributed algorithms also adjust communication frequency based on communication cost and training progress~\cite{wang2019adaptive}. 
However, these approaches optimize training-side or resource-allocation choices in response to observed system conditions, rather than jointly optimizing \mbox{both control planes.}

\bibliographystyle{ACM-Reference-Format}
\bibliography{refs}

\end{document}